\documentclass[twocolumn,prl,superscriptaddress,floatfix]{revtex4}
\usepackage{graphicx}
\usepackage{xcolor}
\usepackage{epsfig}
\usepackage{rotating}
\usepackage{amsmath}
\usepackage{amsfonts}
\usepackage{amssymb}
\usepackage{enumerate}
\usepackage{longtable}
\usepackage{appendix}
\usepackage{dcolumn}% Align table columns on decimal point
\usepackage{bm}
\DeclareMathOperator{\Tr}{Tr}
\usepackage{hyperref}
\usepackage{ulem}

\begin{document}

\title{On the Divergence of the Ferromagnetic Susceptibility in the SU(N) Nagaoka-Thouless Ferromagnet}

\author{Rajiv R. P. Singh}
\affiliation{Department of Physics, University of California Davis, 
CA 95616, USA}

\author{Jaan Oitmaa}
\affiliation{School of Physics, The University of New South Wales,
Sydney 2052, Australia}

\date{\rm\today}

\begin{abstract}

Using finite temperature strong coupling expansions for the SU(N) Hubbard Model, we calculate the thermodynamic properties of the model in the infinite-$U$ limit for arbitrary density $0\leq \rho \leq 1$ and all $N$. We express the ferromagnetic susceptibility of the model as a Curie term plus a $\Delta \chi$, an excess susceptibility above the Curie-behavior. We show that, on a bipartite lattice, graph by graph the contributions to $\Delta \chi$ are non-negative in the limit that the hole density $\delta=1-\rho$ goes to zero. By summing the contributions from all graphs consisting  of closed loops we find that the low hole-density ferromagnetic susceptibility diverges exponentially as $\exp{\Delta /T}$ as $T \to 0$ in two and higher dimensions. This demonstrates that Nagaoka-Thouless ferromagnetic state exists as a thermodynamic state of matter at low enough density of holes and sufficiently low temperatures.  The constant $\Delta$ scales with the SU(N) parameter $N$ as $1/N$ implying that ferromagnetism is gradually weakened with increasing $N$ as the characteristic temperature scale for ferromagnetic order goes down.

\end{abstract}

%\pacs{74.70.-b,75.10.Jm,75.40.Gb,75.30.Ds}

\maketitle

\section{Introduction}

The Hubbard model \cite{Hubbard,kanamori,gutzwiller} is a central model for describing the behavior of electrons in solid state systems and has had a huge impact in our understanding of condensed matter physics \cite{Mattis, Auerbach, Lieb}. Strong correlations, arising from the on-site repulsion in the Hubbard model, can be used to understand many basic solid state phenomena including metal-insulator transitions, antiferromagnetism, superconductivity, spin-liquids and itinerant ferromagnetism.

Nagaoka-Thouless Ferromagnetism is a classic problem in itinerant magnetism \cite{tasaki}. Nagaoka \cite{Nagaoka} and Thouless \cite{Thouless} independently showed that when the Hubbard repulsion $U$ is large enough, a single hole introduced into a system with one-particle per site, polarizes the system around the hole. There have been several variational and numerical studies \cite{shastry, becca, vollhardt, young, liu} of Nagaoka-Thouless ferromagnetism, especially in the ground state of the system. At finite temperatures, rigorous mathematical arguments have been made to show that magnetization in a field exceeds the pure paramagnetic value \cite{aizenman,miyao} and Dynamical Mean-Field Theory (DMFT) \cite{DMFT} was used to obtain a phase diagram in the density-temperature plane.

%While the Hubbard model \cite{Hubbard,kanamori,gutzwiller} is only an approximate model for solid state systems, it has had a huge impact in our understanding of condensed matter physics \cite{Mattis, Auerbach, Lieb}. Strong correlations, arising from the on-site repulsion in the Hubbard model, can be used to understand many basic solid state phenomena including metal-insulator transitions, antiferromagnetism, superconductivity, spin-liquids and itinerant ferromagnetism. 

The cold atomic gases in optical lattices provide a new motivation for study of the Hubbard model \cite{greiner,brown,mitra,cheuk,rosch,gross}.
 In these systems, it is possible to build an ensemble that is well described by the Hubbard model and where the microscopic parameters such as $U$ and $t$ can be controlled and a priori well understood. Furthermore, cold atomic gases allow one to change the number of fermion species from two to a larger $N$ and thus study the Hubbard model with SU(N) symmetry \cite{bloch,honerkamp,taie,lorenzo2,lorenzo,padilla}
 for different values of $N$.

Finite temperature strong-coupling expansion is a natural way to address the magnetic behavior of the Hubbard model at finite temperatures, at various hole densities, in the thermodynamic limit \cite{oitmaa-book, oitmaa2, ten-haaf}. These expansions can be developed in the grand canonical ensemble at fixed fugacity $\zeta=\exp{\beta \mu}$ in powers of $\beta t$, $w=\exp{-\beta U}$, and $1/\beta U$. After changing variables from fugacity to particle density $\rho$, one can obtain temperature dependent thermodynamic properties at various densities.
For $U$ of order or larger than the bandwidth they allow one to relate the thermodynamics of the Hubbard model at low temperatures to a generalized Heisenberg or t-J model \cite{SUN,macdonald,delannoy,mila}. The expansions simplify in the limit $U\to \infty$, in which case many terms can be set to zero and can be used to study the problem of Nagaoka-Thouless ferromagnetism. 

The first few terms of the expansion suffice to give an accurate numerical description of the thermodynamic properties of the model at temperatures larger than the hopping parameter $t$. And, as shown previously for the SU(2) t-J model \cite{putikka,glenister,pryadko}, series extrapolation methods allow one to go to much lower temperatures. But, a numerical extrapolation is difficult to control reliably down to $T=0$. Here, we are interested in the entire temperature range $0<T<\infty$. We show that, close to half filling, i.e. in the limit $\delta=1-\rho$ going to zero, the thermodynamic uniform magnetic susceptibility can be computed all the way to $T\to 0$ by summing over the loop graphs in each order of perturbation theory. This calculation provides a lower bound for the susceptibility and leads to a function which diverges exponentially to infinity as $\exp{\Delta /T}$ as the temperature goes to zero. This shows that, for large enough $U$, the Nagaoka-Thouless ferromagnetic behavior is a thermodynamic phenomena at low density of holes and low enough temperatures. These results are true for any $N>1$ of the SU(N) models \cite{katsura} and in any dimension greater than one. However, the constant $\Delta$ scales as $1/N$, that is the temperature scale for the transition goes down as $N$ increases. For the SU(2) case, our results are in agreement with DMFT which also found that the transition temperature goes to zero as $\delta \to 0$ \cite{DMFT}.

\section{Model and Methods} 

The SU(N) Hubbard model is defined by a Hamiltonian $H=H_0+V$, where the unperturbed Hamiltonian $H_0$ is an on-site term:
\begin{equation}
   H_0= U \sum_i {n_i (n_i-1)\over 2} -\mu \sum_i n_i -h \sum_i (n_{1i} -\frac{n_i}{N}),
\end{equation}
with $n_i$ the total number operator for particles on site $i$ and $\mu$ is the chemical potential. The last term $h$ is a spin-polarizing field that lowers the energy when the particle is in the first spin state $n_{1i}=1$ and raises it for all other states and has an overall zero trace. The perturbation $V$ is the hopping term:
\begin{equation}
    V=-t\sum_{<i,j>}\sum_{\alpha=1}^N (C_{i,\alpha}^\dagger C_{j,\alpha} + h.c.),
\end{equation}
where the sum $<i,j>$ runs over nearest-neighbor pairs of sites of a lattice and the sum over $\alpha$ runs over the $N$ species of Fermions. The total number of fermions of each species is a constant of motion. Thus both the chemical potential and field terms commute with the rest of the Hamiltonian.

Using the formalism of thermodynamic perturbation theory \cite{oitmaa-book, oitmaa2},the logarithm of the grand partition function, per site, can be expended as
\begin{equation} \label{Eq_pert}
    \begin{split}
 \ln{Z} = &\ln{z} +
    \sum_{r=1}^\infty \int_0^\beta d\tau_1 
    \int_0^{\tau_1} d\tau_2 \ldots \int_0^{\tau_{r-1}} d\tau_r \\
    &<\Tilde{V}(\tau_1)\ldots\Tilde{V}(\tau_r)>_N
    \end{split}
\end{equation}
where $z$ is the single-site partition function, 
\begin{equation}
    \Tilde{V} =e^{\tau H_0} V e^{-\tau H_0},
\end{equation}
and,
\begin{equation}
    <X> = \Tr{e^{-\beta H_0}X}/\Tr{e^{-\beta H_0}}.
\end{equation}

In each order, the terms in the expansion can be expressed in terms of various graphs on the lattice as:
\begin{equation}
    \ln{Z}=\ln{z}+\sum_{G} \ L_G\ z^{-N_s} (\beta t)^{N_b} X_G,
\end{equation}
In the expression, the graph $G$ has $N_s$ sites and $N_b$ bonds. $L_G$ is the lattice constant of the graph defined as the extensive part of the graph count, per lattice site. The weight-factor $X_G$ is the reduced contribution of the graph obtained from an evaluation of the traces which depends on $\beta t$, $\beta U$, fugacity $\zeta$, field $h$ and $N$.

In the $U\to \infty$ limit, no double occupancy is allowed and the weight-factor for a graph with $N_s$-sites reduces to
\begin{equation}
    X_G=\sum_{n=1}^{N_s-1} x_n^G \zeta^n.
    \label{xg}
\end{equation}
Here $x_n^G$ is a polynomial in the SU(N) parameter $N$ of order $n$.

From the partition function, the particle density (per site) can be obtained via the relation
\begin{equation}
    \rho=\zeta{\partial \over \partial \zeta} \ln{Z}.
\end{equation}
%This relation needs to be solved to obtain $\zeta$ or $\mu$ as a function of $\rho$ and $\beta$, which then allows one to obtain various properties at fixed particle density.
Thermodynamic quantities such as Internal energy per site, $e$, and entropy per site, $s$, are obtained using the relations
\begin{equation}
    e = -({\partial \over \partial \beta} \ln{Z})_\zeta,
\label{E}
\end{equation}
and
\begin{equation}
    s =-\beta({\partial \over \partial \beta}\ln{Z})_\zeta -\rho \ln{\zeta} + \ln{Z}.
    \label{S}
\end{equation}
The ferromagnetic susceptibility per site is defined by the second derivative of $\ln{Z}$ with respect to the spin-polarizing field $h$. It is given by
\begin{equation}
    \chi=\frac{1}{\beta}\frac{\partial^2}{\partial h^2}
    \ln{Z}
\end{equation}
The field term in the Hamiltonian is defined solely for calculating the susceptibility. Otherwise, we will restrict all calculations to $h=0$.

\section{Single-site Term and Series Expansions}

In the limit of $U\to\infty$ the single-site partition function to order $h^2$ becomes $z= z_0 + h^2 z_1$, where,
\begin{equation}
    z_0= 1+N\zeta 
\end{equation}
and
\begin{equation}
    z_1 = \frac{\beta^2 \zeta}{2} \frac{N-1}{N}
\end{equation}
For all calculations other than the susceptibility, we can set $h=0$. 
In zeroth order the particle density is given by
\begin{equation}
    \rho_0={N\zeta \over 1 + N\zeta}.
\end{equation}
%This can be inverted to give
%\begin{equation}
%    N\zeta = {\rho_0 \over 1-\rho_0}.
%\end{equation}
The zeroth order susceptibility per site is given by
 \begin{equation}
     \chi=\frac{2}{\beta}\frac{z_1}{z_0}
     =(\frac{N-1}{N^2}) \beta \rho.
 \end{equation}
 This is a Curie law as no double occupancy means we have local moments at all temperatures.
 
 We will define excess susceptibility, over and above the Curie-law as
 \begin{equation}
     \chi=C \frac{\rho}{T} + \Delta\chi,
     \label{chi}
 \end{equation}
 with Curie constant $C$ equal to $\frac{N-1}{N^2}$. Note that in this equation the density is the full density not the bare density obtained in zeroth order. Our goal is to calculate $\Delta \chi$.
 
In our studies, we will restrict ourselves to bipartite lattices. All the graphs that contribute to the zero-field partition function on a bipartite lattice to eighth order together with their weights for arbitrary $N$ are given in Supplementary materials.

\begin{figure}[hbpt]
\includegraphics[width=1\columnwidth]{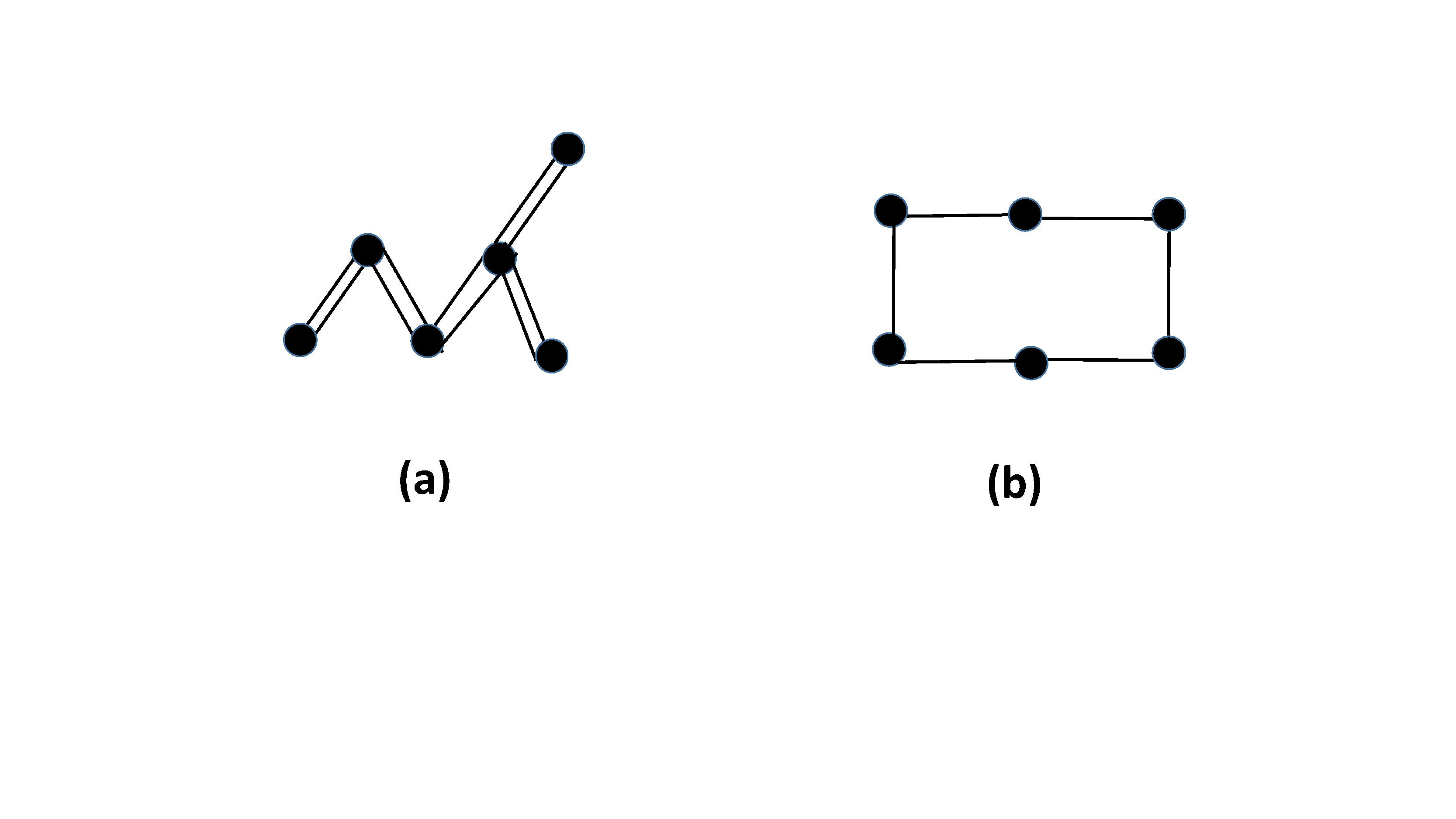}
\caption{Two classes of graphs that contribute to the susceptibility expansion. (a) Tree graphs, with no closed loops. Every bond must be doubled in order to have a non-zero trace contribution. In the single hole sector, each fermion moves back and forth as the holes moves around the graph. Every spin can be independently of any spin species. Such graphs only contribute to the Curie-law and do not contribute to the excess susceptibility at all. (b) Graphs consisting of closed loops. In the single hole sector, as the hole traverses the loop, each spin moves to its neighboring position. In order to have a non-zero trace contribution, all fermions must be of the same species. Thus these graphs have a maximum relative contribution to the excess susceptibility.}
\label{graphs}
\end{figure}

Near one-particle per site all properties can be expanded in powers of the hole density $\delta_0$,
given by
\begin{equation}
    \delta_0 = 1-\rho_0 = \frac{1}{1+N \zeta}=\frac{1}{z_0},
    \label{delta0}
\end{equation}
For the Nagaoka-Thouless problem, we are interested in the limit $\delta=1-\rho$ going to zero. Thus, we will keep terms linear in $\delta_0$ and drop all terms proportional to higher powers of $\delta_0$. These linear in $\delta_0$ terms come from exactly one hole in each cluster. We should note that this does not mean we are looking at a single hole in the thermodynamic system. Our formalism implies that we are studying the limit of low hole density as similar behavior will be happening independently all over the system.
  
In the large $U$ limit, the weight of a graph $X_G$ is a polynomial in $\zeta$ of order $N_s-1$ , where $N_s$ is number of sites in the cluster. The restriction to lowest power of $\delta_0$ reduces the weight factor for a graph to:
\begin{equation}
    X_G= x_G \zeta^{N_s-1}.
    \label{xg2}
\end{equation}
The coefficients  $x_G$, which depend on $N$ and the field $h$, turn out to be always positive as can be seen from the explicit calculations to eighth order in the supplementary materials. These terms correspond either to a single hole moving back and forth on a tree like graph with no closed loops or a single hole moving in  closed loops. In both cases they are positive. We will see that this will lead us to the result that the contribution to excess susceptibility from every graph is non-negative. This means that even a partial summation of graphs is a lower bound on the excess susceptibility.

In this limit, the relation between the full density function and fugacity becomes
\begin{equation}
    \rho=\rho_0 + \sum_G L_G(\beta t)^{N_b} (N_s-1-N_s\rho_0)\ \frac{x_G \zeta^{N_s-1}}{z_0^{N_s}}
\end{equation}
The excess susceptibility is given by
\begin{equation}
    \Delta\chi=\sum_G L_G \ (\beta t)^{N_b} (C_{1G}-C_{2G}),
\end{equation}
where
\begin{equation}
    C_{1G} = \frac{1}{\beta}\frac{\partial^2}{\partial h^2}
    (\frac{X_G}{z^{N_s}}),
\end{equation}
and
\begin{equation}
    C_{2G}=(N_s-1-N_s\rho_0)\frac{C}{T}\frac{X_G}{z^{N_s}}.
\end{equation}

For tree graphs, with no closed loops (see Fig.~1), the coefficient $x_G$ in zero field is proportional to $N^{N_s-1}$. This reflects the fact that in the absence of closed loops every spin can independently be of any species. For these graphs the contribution to excess susceptibility vanishes identically. Physically this is a reflection of the fact that susceptibility of independent spins is already contained in the Curie law. Thus, we only need to consider those weights where the power of $N$ in a graph with $N_s$ sites is less than $N_s-1$. In all these terms at least some of the spins are constrained to be of the same species. Even the smallest such constraint can be shown to lead to a positive contribution to the susceptibility.

At the other extreme are those terms where zero-field $x_G$ scales linearly with $N$. This implies that every spin in the graph must be of the same species to contribute to a non-zero trace. An example is a graph consisting of a single closed loop (See Fig.~1). It must have this behavior. As a hole traverses around the loop, every Fermion in the loop moves to its neighboring position and hence must be of the same species as its neighbor to contribute to the trace. These graphs contribute maximally to the excess susceptibility. It can be shown that for a single loop of length $l$ the zero-field weight-factor is
\begin{equation}
    X_G=\frac{2l}{l!}N \ \zeta^{l-1}
\end{equation}
The excess susceptibility contribution from this graph can be shown to be
\begin{equation}
  \frac{C}{T} \frac{(\beta t)^{l}}{l!}2 l (l-1)(l-2) N \frac{\zeta^{l-1}}{z_0^l}.
\end{equation}
Expressing this in terms of $\rho_0$ gives
\begin{equation}
  \frac{C}{T} \frac{(\beta t)^{l}}{l!}2 l (l-1)(l-2) \frac{\rho_0^{l-1}\delta_0}{N^{N_s-2}}.
\end{equation}
It is well known that for large $l$, the number of polygons of even length $l$ embedded in a bipartite lattice scale as \cite{polygon}
\begin{equation}
    p_l= A \mu_p^l\  l^{a-3},
\end{equation}
where the constant $\mu_p$ called the connectivity constant is known approximately for most lattices \cite{polygon2}. Ignoring the weak dependence on the exponent $a$ which will only affect the prefactor, the contributions of polygons can be summed to obtain an excess susceptibility of
\begin{equation}
    \Delta \chi \propto N^2 \frac{\delta_0}{\rho_0} \frac{C}{T}\exp{ \frac{\Delta}{T}},
\end{equation}
with $\Delta=\frac{t\rho_0\mu_p}{N}.$
Ignoring the slowly varying prefactor, this shows that the excess susceptibility diverges exponentially as $T\to 0$. We believe the primary role of the additional terms not included in this summation is to decorate these graphs and renormalize the bare density $\rho_0$ to the full density $\rho$.

This result provides a lower bound to the magnetic susceptibility and implies that the Nagaoka-Thouless ferromagnet is a thermodynamic phase of matter for low hole density and low enough temperatures. The characteristic temperature scale at which the susceptibility becomes exponentially large is inversely proportional to the SU(N) parameter $N$. Thus, the tendency for ferromagnetism gradually weakens with increase in $N$. These results are in agreement with the earlier dynamical mean-field theory results for the SU(2) case in that the ferromagnetic phase boundary was found to go to zero temperature as the hole density goes to zero \cite{DMFT}. They are also in agreement with mathematical arguments that show the existence of finite magnetization in a field that exceeds the paramagnetic value at any temperature \cite{miyao}.

The extension of these results to finite $U/t$ and finite hole doping can be done numerically as was done for the SU(2) t-J models some time ago \cite{putikka,glenister,pryadko,yedidia}. Those studies show that at small enough $J/t$ and close to half filling the peak in the magnetic susceptibility shifts to $q=0$. However, from a small number of terms in the expansion it is more difficult to rigorously establish the divergence of the susceptibility.

%but the $h^2$ dependence come both from the numerator which gives a positive contribution to the susceptibility and powers of $z$ in the denominator which give a negative contribution to the susceptibility. The coefficient of $\zeta^{n-1}$ is a polynomial in the SU(N) variable $N$. The dominant terms to the susceptibility come from the first power of $N$, which corresponds to all fermions belonging to the same spin species. The numerator terms in this case scale in order $n$ as $n^2$ which far exceeds the negative term coming from the denominator, which scales only linearly with $n$. Thus these dominant terms are all positive.

\section{Discussions and Conclusions}
In this paper we have used finite temperature strong coupling expansions for the Hubbard model to revisit the problem of Nagaoka-Thouless ferromagnetism at large $U$ and small hole doping near one-particle per site. We have shown that the ferromagnetic susceptibility of the system diverges exponentially as $\exp{\Delta/T}$ as $T\to 0$. Thus at sufficiently low temperatures, the system must turn ferromagnetic.

While the Hubbard model is an approximate model for solid state systems, it can be a well characterized and accurate model in cold atomic gases in optical lattices. Furthermore, in these systems, the Hubbard parameters can be tuned by lasers and the number of fermion species can be made larger than $2$ and the system can have SU(N) symmetry. We have shown that Nagaoka-Thouless ferromagnetism is present for all $N$ and is only weakened gradually with increase in N. Fundamentally, this ferromagnetism arises from the fact that as a hole traverses a closed loop, non-zero trace arises only if all the spins belong to the same species. This combined with positive trace on loops of bipartite lattices implies an exponentially divergent susceptibility as $T \to 0$. 

We hope our work would stimulate further experimental search for Nagaoka-Thouless ferromagnetism and measurements of the temperature dependence of the susceptibility at low hole densities in solid state and cold atom systems.

{Acknowledgement:} This work is supported in part by the US National Science Foundation grant DMR-1855111.

\end{document}